\documentclass[3p,twocolumn]{elsarticle}

\usepackage{amsmath}
\usepackage{amssymb}
\usepackage{graphicx}
\usepackage{color}
\usepackage{slashed}

\newcommand{\GeV}{\rm{\,GeV}}
\newcommand{\TeV}{\rm{\,TeV}}

\author[tri,nic]{E. Gabrielli}
\author[nic]{K. Kannike}
\author[rom]{B. Mele}
\author[nic,ut]{M. Raidal}
\author[nic]{C. Spethmann}
\author[nic]{H. Veerm\"ae}

\address[tri]{Dipart. di Fisica Teorica, Universit\`a di 
Trieste, Strada Costiera 11, I-34151 Trieste, Italy and 
INFN, Sezione di Trieste, Via Valerio 2, I-34127 Trieste, Italy}
\address[nic]{NICPB, R\"avala 10, Tallinn 10143, Estonia}
\address[rom]{INFN, Sezione di Roma, c/o Dipart. di Fisica, Universit\`a di Roma ``La Sapienza", \\ Piazzale Aldo Moro 2, I-00185 Rome, Italy}
\address[ut]{Institute of Physics, University of Tartu, Estonia}
\title{A SUSY Inspired Simplified Model for the 750 GeV Diphoton Excess}

\begin{document}

\begin{abstract}
The evidence for a new singlet scalar particle from the 750~GeV diphoton excess, and the absence of any other signal of new physics at the LHC so far,
suggest the existence of new coloured scalars. To study this possibility, we propose a supersymmetry inspired simplified model, extending the Standard Model  
with a singlet scalar and with heavy scalar fields carrying both colour and electric  charges -- the `squarks'. To allow the latter to decay, and to generate the dark matter 
of the Universe, we also add a neutral fermion to the particle content. We show that this model provides a two-parameter fit to the observed diphoton excess consistently 
with cosmology, while the allowed parameter space is bounded by the consistency of the model. In the context of our simplified model this implies the existence of other
supersymmetric particles accessible at the LHC, rendering this scenario falsifiable. If this excess persists, it will imply a paradigm shift 
in assessing supersymmetry breaking and the role of scalars in low scale physics.
\end{abstract}

\maketitle

%%%%%%%%%%%%%%%%%%%%%%%%%%%%%%%%%%%%%%%%%%%%%%%%%%%%%%%%%%%%%%%%
\section{Introduction}

The discovery of the Higgs boson at the LHC~\cite{Aad:2012tfa,Chatrchyan:2012xdj} completed the observation of all fundamental degrees of freedom predicted by the Standard Model (SM)  of particle  interactions.
Nevertheless, it is widely believed that the SM suffers from a series of shortcomings, related to the stability of the electroweak scale and the absence of a candidate for the dark matter (DM) of the Universe, for instance. Solutions to these problems require extending the present theoretical framework to include new degrees of freedom, possibly relevant at 
the energy scales probed by colliders in the near future.

The LHC Run 2 at 13 TeV collision energy provides the potential to probe physics at shorter distances compared to LHC Run 1 at 7~TeV and 8~TeV. 
The exploration has just started with about 4 fb$^{-1}$ of integrated luminosity delivered to the ATLAS and CMS experiments, beginning in June 2015. 
Searching for two-particle resonances is an especially adequate way to look for new physics manifestations when new thresholds in collision energies are reached. 
Both  ATLAS and CMS are presently analysing the new data sets, and trying to get  the most out of the very first run at 13 TeV.

In a recent CERN seminar  both ATLAS~\cite{ATLAS} and CMS~\cite{CMS} presented a photon pair excess with an invariant mass at about 750 GeV, with a {\it local} significance varying (depending on the narrow- or wide-width assumption) in the range 2.6 to 3.9~$\sigma$. 
The signal also exhibits some compatibility with the photon-pair studies of Run 1 data by both ATLAS and CMS. 
Assuming that the observed diphoton excess is due to a new resonance, CMS provides a combination of  Run 1 plus Run 2 data for its production cross section
times branching fraction into photons to be $4.5\pm 1.9$~fb~\cite{CMS}. The corresponding ATLAS result for 13~TeV was estimated to be 
$10.6\pm2.9$~fb~\cite{DiChiara:2015vdm}.
While further data will be needed in order to clarify whether the observed excess is robust, it is exciting to  assume that the di-photon excess is really pointing to 
the existence of new physics below a scale of 1~TeV, and to try to determine which kind of SM extension can predict such an effect. 
Presently, no  anomaly in any other final state has been detected~\cite{ATLAS,CMS}, which severely restricts any realistic explanation of the excess.

The most natural interpretation of the observed diphoton excess is due to the decays of a singlet scalar $S$ into photons, 
$S\to\gamma\gamma$~\cite{DiChiara:2015vdm,Franceschini:2015kwy,Knapen:2015dap,Buttazzo:2015txu,Backovic:2015fnp,Mambrini:2015wyu,Gupta:2015zzs,Ellis:2015oso,McDermott:2015sck,Dutta:2015wqh,Cao:2015pto,Kobakhidze:2015ldh,Martinez:2015kmn,Bian:2015kjt,Chakrabortty:2015hff,Falkowski:2015swt,Bai:2015nbs}.\footnote{Solutions with pseudoscalars have also been considered in~\cite{DiChiara:2015vdm,Pilaftsis:2015ycr,Low:2015qep,Higaki:2015jag,Molinaro:2015cwg,Becirevic:2015fmu}.}  
The existence of light scalars much below the cut-off scale of the SM (such as the Planck scale) requires some mechanism to protect their masses
against radiative corrections from the cut-off scale. By far the most popular solution to the hierarchy problem is supersymmetry.
However, in recent years supersymmetry has lost some of its appeal because of the severe experimental constraints from the LHC~\cite{ATLAS,CMS}.

In the context of the diphoton excess the conventional supersymmetric models, such as the Minimal Supersymmetric Standard Model (MSSM), have several shortcomings.
Firstly, the excess cannot be explained within the MSSM alone~\cite{DiChiara:2015vdm,Buttazzo:2015txu,Angelescu:2015uiz,Gupta:2015zzs},
and the framework must be extended  to accommodate the new signal. This points into the direction of the Next-to-Minimal Supersymmetric Standard Model (NMSSM) 
rather than the MSSM. Secondly, most of the diphoton excess studies so far have assumed the existence of heavy coloured vectorlike fermions that, at one loop, induce singlet scalar couplings to gluons and photons~\cite{Franceschini:2015kwy,Knapen:2015dap,Pilaftsis:2015ycr,Buttazzo:2015txu,Angelescu:2015uiz,Gupta:2015zzs,Ellis:2015oso,McDermott:2015sck,Kobakhidze:2015ldh,Martinez:2015kmn,No:2015bsn,Chao:2015ttq,Fichet:2015vvy,Curtin:2015jcv,Falkowski:2015swt}. 
Such coloured fermions, however,  do not exist in the particle content of any supersymmetric extension
of the SM. In addition, new coloured fermions are severely constrained by LHC searches and must be very heavy~\cite{ATLAS,CMS}. 
Therefore, extending the non-minimal supersymmetric
 models further with charged and coloured vectorlike fermions implies that the model that is supersymmetrised is not the SM.  The model becomes unnecessarily 
 complicated without any obvious  need for these specific new particles.
 
 In this work we argue that the diphoton excess hints at the existence of relatively light coloured and charged scalars. 
 First, these particles, {\it the squarks}, do exist in any supersymmetric extension of the SM and there is no need to 
 extend the model with new {\it ad hoc} particles. 
 Second, the LHC constraints on coloured scalar masses  are much less stringent than on coloured fermions, such as gluinos.\footnote{The only exception are strongly coupled scalar diquarks~\cite{Ma:1998pi}, exotic scalars coupled to two valence quarks, that should produce quark-quark resonances at the LHC and
 which masses are, therefore, constrained to be above 6~TeV~\cite{Khachatryan:2015dcf} The models with coloured scalars in the loop
 presented in Ref.~\cite{Knapen:2015dap} are based on diquarks that are not superpartners of quarks. Also, their model does not contain dark matter candidates.}
 These arguments allow for relatively light squarks in the loops generating $gg\to S\to\gamma\gamma$ that are potentially observable 
 at the LHC in coming years, rendering this scenario directly testable.
 Third, one of the favourable feature of supersymmetric models is the existence of dark matter (DM) that comes for free as the lightest neutral superpartner of 
 gauge bosons and scalars. We use these arguments to address the LHC diphoton excess.

 Motivated by the above mentioned good features of supersymmetric theories, we propose a supersymmetry inspired
 simplified model that is able to explain the diphoton excess consistently with all other LHC results and with the existence of DM.
 Although minimal by construction, and therefore not supersymmetric by itself, this model uses the particle content of the NMSSM 
 and can be embedded into the latter.\footnote{Alternatively, the singlet could be
 a sgoldstino~\cite{Petersson:2015mkr,Bellazzini:2015nxw,Demidov:2015zqn}, the superpartner of supersymmetry breaking goldstone.}
 Therefore, the mass spectrum of this type of supersymmetric models must be very different from the ones predicted by simple supersymmetry
 breaking scenarios.

 We study this effective model carefully and show that the requirement of a physical, charge and colour conserving vacuum restricts the allowed mass parameters to be constrained from above, rendering the model testable or falsifiable by collider experiments. In the context of the simplified model this statement means that the effective theory
 breaks down and the new supersymmetric degrees of freedom must appear to cure the model. Therefore, if verified, our framework {\it predicts} the existence
 of new supersymmetric particles at the reach of next collider runs. 
Thus the di-photon excess may change our present understanding of the supersymmetry breaking patterns and the role
 of scalars in supersymmetric models. 
 %Relaxing our prejudices about the supersymmetric particle spectrum would allow us to discover it at the LHC. 

\begin{figure}[t]
\begin{center}
\includegraphics{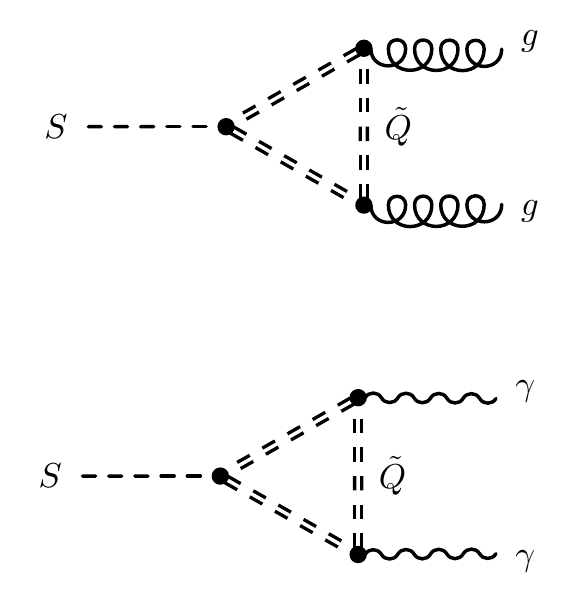}
\caption{Leading order contributions to the main decay modes of $S$.}
\label{fig:Sdec}
\end{center}
\end{figure}

%%%%%%%%%%%%%%%%%%%%%%%%%%%%%%%%%%%%%%%%%%%%%%%%%%%%%%%%%%%%%%%%
\section{The SUSY Inspired Simplified Model}

We construct a supersymmetry inspired simplified model that produces a narrow scalar resonance  $gg \to S \to \gamma \gamma$. This resonance is necessarily a singlet under the SM gauge group. As shown in Fig.~\ref{fig:Sdec}, its interactions with photons and gluons are therefore induced at loop level by another scalar field $\tilde{Q}$ that is coloured and carries  hypercharge, which we assume to be $q_{\tilde{Q}} = 2/3$. $\tilde{Q}$ is therefore identical to the well known right-handed up-type squark that transforms in the fundamental representation of $SU(3)$ but is a singlet under $SU(2)$. 
To avoid any conflict with LHC phenomenology and  cosmological and astrophysical observations, the squark $\tilde{Q}$ is required to be unstable. 
As in the supersymmetric extension of the SM, we take it to decay into a quark and a neutralino-like fermion $\chi_{0}$, which is the dark matter candidate in our scenario.

Thus we consider a minimal extension of the SM with the real singlet scalar field $S$, three generations of `squarks' $\tilde{Q_{i}}$ and the `neutralino' $\chi_{0}$. Obviously, the model with this particle-content is by itself not supersymmetric, but requires embedding into a supersymmetric theory. The general Lagrangian for the given particle sector contains the following terms
\begin{align}
	{\cal L}_{\rm kin} &	= |D_{\mu} \tilde{Q}_{i}|^{2} + \frac{1}{2}(\partial_{\mu} S) (\partial^{\mu} S)
	\\
& - 	M^2_{\tilde{Q}} |\tilde{Q_{i}}|^{2}	- \frac{1}{2} M^2_{S0} S^{2} 
	+ \frac{1}{2}	\bar{\chi}_{0}(\slashed{\partial} - m_{\chi_{0}})\chi_{0}, \notag
	\\
	{\cal L}_{\rm dec}
&	= 	\frac{1}{2} y_{\chi} S \chi^{T}_{0}\chi_{0}
	+	(y_{i} \tilde{Q_{i}}^{\dagger}\chi^{T}_{0} U_{iR} + \text{h.c.}), 
	\\ 
	\label{Lag_scalar}
	{\cal L}_{\rm scalar}
&	= 
	- \mu_{\tilde{Q}} S |\tilde{Q_{i}}|^{2}
	- \frac{\mu_{S}}{3} S^{3} 
	- \frac{\lambda_{S}}{4} S^{4}
       \\   
&
	- \lambda_{S\tilde{Q}} S^{2} |\tilde{Q_{i}}|^{2} 
	- \lambda_{\tilde{Q}} |\tilde{Q_{i}}|^{2} |\tilde{Q_{j}}|^{2} \notag
	\\
&	- \lambda'_{\tilde{Q}} (\tilde{Q}_{i}^{\dag} \tilde{Q}_{j}) (\tilde{Q}_{j}^{\dag} \tilde{Q}_{i}),  \notag
	\\    
	{\cal L}_{H}
&	= 
	- (\mu_{H} S + \lambda_{HS} S^{2}) H^{\dag} H \\
&	- \lambda_{\tilde{HQ}} (\tilde{Q_{i}}^{\dag} \tilde{Q_{i}}) (H^{\dag} H), \notag
\end{align}
with the covariant derivative $D_{\mu}=\partial_{\mu}+g_sT^a G^a_{\mu} +e q_{\tilde{Q}} A_{\mu}$,  where $G^a_{\mu}$ is the gluon and $A_{\mu}$ is the photon field, and we sum over the generation indices $i$ for $\tilde{Q}_{i}$. We assume a flavour symmetry to forbid any other terms involving $\tilde{Q}_{i}$. Eq.~\eqref{Lag_scalar} contains the interactions among the two scalars, most importantly the first term with the coupling $\mu_{\tilde{Q}}$ which has dimensions of mass.

We require that $M_{\chi}, M_{\tilde{Q}} > M_{S}/2 $ to forbid tree level decays of the $S$ resonance. Also, instability of $\tilde{Q}$ dictates that $M_{\tilde{Q}} > M_{\chi}$. This choice has the benefit of providing a dark matter candidate  -- the neutralino $\chi_{0}$. 
It has recently been shown~\cite{Backovic:2015fnp,Mambrini:2015wyu}
 that in such a setup the observed amount of DM can be produced from thermal freeze-out analogously to the MSSM.
Thus the DM in our scenario is a thermal relic in the form of the stable neutralino $\chi^0.$ To satisfy the observation $\Omega_{\rm DM} h\sim 0.1,$
the neutralino mass must be ${\cal O}(300)$~GeV~\cite{Backovic:2015fnp,Mambrini:2015wyu}
, implying a somewhat compressed spectrum.
The latter implies that the model is not severely constrained by the LHC searches for squark pair production for the final state of two jets and missing energy~\cite{Aad:2015iea}.

As we have already commented, this model does not fit into the MSSM but requires some extended supersymmetric model, the NMSSM being 
the simplest of them. We note that the mass spectrum of such a model must  feature light scalars while the gluino must be heavy to comply with the LHC bound.
Since our study is phenomenological, we just assume this supersymmetry breaking pattern.

%%%%%%%%%%%%%%%%%%%%%%%%%%%%%%%%%%%%%%%%%%%%%%%%%%%%%%%%%%%%%%%%
\section{Conditions for a Physical Vacuum} 

We consider the conditions for the vacuum of the model  not to break colour and electric charge. We need to ensure the following:
\begin{enumerate}
\item The potential is bounded from below in the limit of large field values.
\item The squarks $\tilde{Q}_{i}$ do not get VEVs, which would break colour and electric charge. The true vacuum should be at $S = 0$ and $\tilde{Q} = 0$, therefore the potential has to be positive everywhere else,\footnote{The Higgs portal couplings are already strongly constrained~\cite{Falkowski:2015swt} and we neglect them for phenomenological reasons.  We also assume that $\mu_{H} \simeq 0$ to prevent a large decay width of $S$ into Higgs bosons.}
\begin{equation}
  V(S \neq 0, \tilde{Q} \neq 0) > 0.
\end{equation}
\item $S$ does not get a VEV: a non-zero VEV for $S$ would shift the mass of $S$ away from $M_{S}$.\footnote{A VEV for $S$ would also generate large contribution to the mass of the squark, which would need to be fine-tuned.}
\end{enumerate}

The potential must be bounded from below in order for a finite minimum of potential energy to exist. In the limit of large field values, we can ignore the dimensionful terms in the scalar potential. The full bounded below conditions can be found via co-positivity constraints on the quartic part of the scalar potential \cite{Kannike:2012pe}:
\begin{align}
& \lambda_{S} > 0, \quad \lambda_{\tilde{Q}} + \theta(-\lambda'_{\tilde{Q}}) \lambda'_{\tilde{Q}} > 0, \quad \lambda_{H} > 0, \\
& \bar{\lambda}_{SQ} \equiv 2 \sqrt{\lambda_{S} [\lambda_{\tilde{Q}} + \theta(-\lambda'_{\tilde{Q}}) \lambda'_{\tilde{Q}}]} + \lambda_{S\tilde{Q}} > 0, \\
& \bar{\lambda}_{HQ} \equiv 2 \sqrt{\lambda_{H} [\lambda_{\tilde{Q}} + \theta(-\lambda'_{\tilde{Q}}) \lambda'_{\tilde{Q}}]} + \lambda_{H\tilde{Q}} > 0, \\
& \bar{\lambda}_{HS} \equiv 2 \sqrt{\lambda_{H} \lambda_{S}} + \lambda_{HS} > 0, \\
& \lambda_{HS} \sqrt{\lambda_{\tilde{Q}} + \theta(-\lambda'_{\tilde{Q}}) \lambda'_{\tilde{Q}}} + \lambda_{H\tilde{Q}} \sqrt{\lambda_{S}} + \lambda_{S\tilde{Q}} \sqrt{\lambda_{H}}
\notag  \\
& + 2 \sqrt{[\lambda_{\tilde{Q}} + \theta(-\lambda'_{\tilde{Q}}) \lambda'_{\tilde{Q}}] \lambda_{S} \lambda_{H}}
\\
& + \sqrt{\bar{\lambda}_{SQ} \bar{\lambda}_{HQ} \bar{\lambda}_{HS}}  > 0,   \notag
\end{align}
where $\theta$ is the Heaviside step function. The conditions can be satisfied by taking $\lambda_{S} \geq 0$, $\lambda_{\tilde{Q}} \geq 0$, $\lambda'_{\tilde{Q}} \geq 0$, $\lambda_{S\tilde{Q}} \geq 0$, $\lambda_{HS} \geq 0$, $\lambda_{H\tilde{Q}} \geq 0$. 
The stationary point equations for the new particles are then
\begin{align}
0 &= \mu_{\tilde{Q}} |\tilde{Q}_{i}|^{2} + (M_{S}^{2} + 2 \lambda_{S\tilde{Q}} |\tilde{Q}_{i}|^{2}) S  \label{eq:min:S} \\
& + \mu_{S} S^{2} + \lambda_{S} S^{3},  \notag \\
 0 &= |\tilde{Q}_{i}| (M_{\tilde{Q}}^{2} + 2 \lambda_{\tilde{Q}} |\tilde{Q}_{i}|^{2} + \mu_{\tilde{Q}} S + \lambda_{S\tilde{Q}} S^{2}). \label{eq:min:Q}
 \end{align}
If $\tilde{Q}_{i} = 0$, we need 
\begin{equation}
  \mu_{S}^{2} < 4 \lambda_{S} M_{S}^{2}
\end{equation}
 for $S$ not to get a VEV. We will take $\mu_{S} \simeq 0$ to get the largest allowed parameter space for the diphoton signal. However, we note that a small but non-zero $\mu_{S}$ could always be generated at two loops.

$S$ and $\tilde{Q}_{i}$ could also get non-zero VEVs simultaneously. We need to forbid this to prevent a coloured vacuum. The forbidden part of the parameter space is found by requiring that the vacua where $S$ and $\tilde{Q}_{i}$ have non-zero VEV -- if they exist -- are local minima of the potential, that is, $V > 0$.

Note that the bound does not depend on the number of flavours, which cancels out in the minimisation equations. Especially, the $\lambda'_{\tilde{Q}}$ term does not affect the result, since its minimal value is zero. Also, it is plausible that the bound will not be weakened by much if the flavor symmetry is abandoned.

\begin{figure}[t]
\begin{center}
\includegraphics[width=0.45\textwidth]{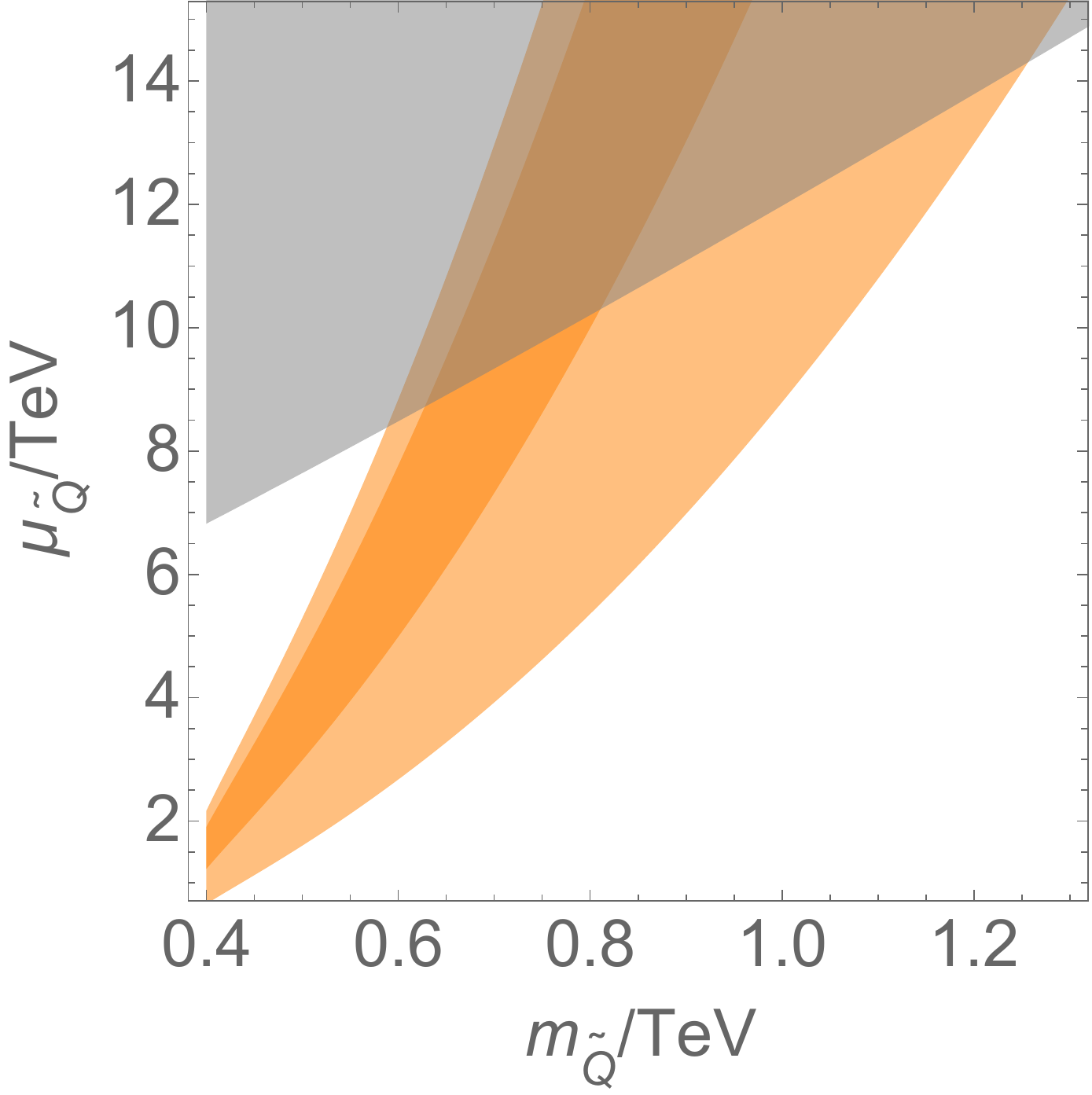} 
\caption{$\sigma(pp \to S) \times \text{BR}(S \to \gamma \gamma)$ at the 13 TeV LHC. The colored regions correspond to 4.5$\pm$1.9 fb (inner region) and 4.5$\pm$ 3.8 fb (outer region) corresponding to the $1\sigma$ and $2\sigma$ regions for $N_f = 3$ degenerate squark generations. The horizontal axis shows the mass of the colored scalar particle $\tilde{Q}$  and the vertical axis the trilinear $S\tilde{Q}\tilde{Q}$ coupling. The grey shaded region is forbidden by the presence of colour symmetry breaking assuming  $\lambda_{\tilde{Q}} = \lambda_{S} = \lambda_{S\tilde{Q}} = 4 \pi$.}
\label{fig:XSecPlot}
\end{center}
\end{figure}

To fit the diphoton signal we need a large $\mu_{\tilde{Q}}$ that tends to destabilise the SM vacuum. This effect can be countered with large quartic couplings. In Fig.~\ref{fig:XSecPlot} we show the forbidden region on $\mu_{\tilde{Q}}$ vs. $M_{\tilde{Q}}^{2}$ plane with gray colour for the least constraining choice $\lambda_{\tilde{Q}} = \lambda_{S} = \lambda_{S\tilde{Q}} = 4 \pi$. In the context of this effective model, that must be embedded into a supersymmetric model, the appearance of non-perturbative
couplings signals the break-down of the effective model. This implies that the supersymmetric particles of the full model must appear below the scale given by this constraint.

%%%%%%%%%%%%%%%%%%%%%%%%%%%%%%%%%%%%%%%%%%%%%%%%%%%%%%%%%%%%%%%%
\section{Event Rates}

We choose the mass of the singlet to be on the resonance, $M_{S} = 750$ GeV. At this energy scale $\alpha_s(M_{S}) = 0.0894(31)$ \cite{Chatrchyan:2013txa}, whereas for $\alpha = 1/137$ we use the zero momentum value.
From the CMS~\cite{CMS} we know that $\sigma (pp \to S \to \gamma \gamma) \simeq 4.5~\text{fb}$.

The partial decay widths of the singlet $S$ into two photons and into two gluons are \cite{Gunion:1989we,Djouadi:2005gj} %Dj p92, p118
\begin{align}\label{eq:partial_G}
  \Gamma (S \to \gamma \gamma) 
  &= \frac{ \alpha^{2} M_{S}^{3}\mu_{\tilde{Q}}^{2}}{1024 \pi^{3}M_{\tilde{Q}}^{4}} N_{f}^{2} N_{c}^{2} q_{\tilde{Q}}^{4} \left| A_{0}(\tau) \right|^{2},
	\\
  \Gamma (S \to g g) 
  &= \frac{ \alpha_{s}^{2} M_{S}^{3} \mu_{\tilde{Q}}^{2}}{512  \pi^{3} M_{\tilde{Q}}^{4}} N_{f}^{2} \left| A_{0}(\tau) \right|^{2},
\end{align}
respectively. $N_c = 3$ denotes the dimension of the representation for the squarks and $N_f$ the number of squark flavors. 
The scalar loop function is given by~\cite{Gunion:1989we} %p26
\begin{equation}
  A_{0}(\tau) = \tau(1 - \tau f(\tau)),
\end{equation}
with $\tau = 4 M_{\tilde{Q}}^{2}/M_{S}^{2}$, and the universal scaling function is
\begin{equation}
 \! f(\tau) = 
  \begin{cases}
    \arcsin^{2} \sqrt{1/\tau}
    & \tau \geq 1 ,
    \\
%   -\frac{1}{4} \left( \log \frac{1+\sqrt{1-\tau}}{1-\sqrt{1-\tau}} - i \pi \right]^{2}
 	-\left( {\rm arccosh} \sqrt{1/\tau} - i \pi/2 \right)^{2}
    & \tau < 1.
  \end{cases}
\end{equation}

The cross section for producing the diphoton signal via the decay of $S$ in the narrow width approximation is
\begin{equation}
  \sigma(gg \to S \to \gamma \gamma) = \sigma (gg \to S) \, \text{BR}(S \to \gamma \gamma),
\end{equation}
where the production cross section is related to the decay width into gluons by
\begin{equation}\label{eq:partonicGGF}
  \sigma(gg \to S) = \frac{\pi^{2}}{8 M_{S}} \Gamma (S \to g g) \delta(\hat s - M_{S}^{2}).
\end{equation}
Taking into account that $\Gamma (S \to \gamma \gamma) \ll \Gamma (S \to g g) \simeq \Gamma_{S}$, the branching ratio reads
\begin{equation}
  \text{BR}(S \to \gamma \gamma) 
  \simeq \frac{1}{2} \frac{\alpha^{2}}{\alpha_{s}^{2}} N_{c}^{2}q_{\tilde{Q}}^{4} 
  \simeq 0.58\%, %\times \left(\frac{q_{\tilde{Q}}}{2/3}\right)^{4}
\end{equation} 
where we used $\alpha_s(M_{S}) \simeq 0.09$, $\alpha \simeq 1/137$ and assumed the up type squark with the charge $q_{\tilde{Q}} = 2/3$. We remark that if the dominant decay mode of $S$ is $S \to g g$ as assumed here, the cross section for the resonant production of diphotons by gluon-gluon fusion is approximately independent of the details of the strong interaction since
\begin{equation}\label{eq:partonicGGFapprox}
	\sigma(gg \to S \to \gamma \gamma) 
	\simeq \frac{\pi^{2}}{8 M_{S}} \Gamma (S \to \gamma\gamma)\delta(\hat s - M_{S}^{2}).
\end{equation}
At the level of precision considered here, we assume that this cancellation also holds if higher orders corrections in $\alpha_s$ are taken into account.

To calculate the $S$ resonance production cross section at the LHC, we integrate Eq.~\eqref{eq:partonicGGFapprox} numerically using the MSTW parton distribution function (pdf) set \cite{Martin:2009iq}
\begin{align}
	\sigma(gg \to S \to \gamma \gamma)  = \frac{\pi^{2}}{8 M_{S}^3} I_{\rm pdf} \Gamma (S \to \gamma\gamma) ,
\end{align}
where $\sqrt{s} = 13\,\TeV$ is the center of mass energy of LHC proton-proton collisions, and
\begin{align}
	I_{\rm pdf} 
	= \int_{M_S^2/s}^1 \; \frac{\mathrm{d}x}{x} \; \bar{g}(x) \; \bar{g}\left(\frac{M_S^2}{sx}\right)
	\approx 5.8,
\end{align}
is the dimensionless pdf integral evaluated at $\sqrt{s} = 13\,\TeV$. Here $g(x, M_S) = \bar{g}(x, M_S)/x$ is the pdf of the gluon at momentum fraction $x$ evaluated at the scale $M_S=750$~GeV. 

To reproduce the observed signal, we find that the partial decay width to photons is
\[ 
\Gamma(S \to \gamma \gamma) \approx  (0.68 \pm 0.28) \mbox{  MeV}.
\]
The parameter space that reproduces the observed decay width for $N_f = 3$ generations is depicted in Fig.~\ref{fig:XSecPlot}. Accounting for unitarity and preserving color and charge symmetries, it follows, that within the $1\sigma$ band the data favors $N_f \geq 2$ generations of light squarks with masses below $M_{\tilde{Q}} \lesssim 800\GeV$ and a relatively large coupling to the scalar $S$ of $\mu_{\tilde{Q}} \gtrsim 2\TeV$. As was also noted in \cite{Gupta:2015zzs} in the context of a different model, we similarly find that the signal can not be reproduced by a single generation of light squarks within $1\sigma$.

The most important result evident in Fig.~\ref{fig:XSecPlot} is that the allowed parameter space of this effective model 
is bounded to a small region by the di-photon excess and by the 
consistency of the effective model. This implies that new particles must be present in Nature at the scale ${\cal O}(1)$~TeV.

%%%%%%%%%%%%%%%%%%%%%%%%%%%%%%%%%%%%%%%%%%%%%%%%%%%%%%%%%%%%%%%%
\section{Effective Field Theory Approach}

We turn to analyse our scenario in terms of the effective Lagrangian approach.
In the case the squark is heavier than the singlet $S$, the latter can acquire effective couplings with photons and gluons by integrating out the squark field. 
For generic squarks  this corresponds to an effective Lagrangian
\begin{equation}
	{\mathcal{L}}_{\rm eff}
	= \frac{1}{\Lambda_{\gamma}} S \,F^{\mu \nu}F_{\mu \nu} + \frac{1}{\Lambda_{G}} S \,G^{a \mu \nu}G^a_{\mu \nu},
\label{Leff}
\end{equation}
where $F_{\mu \nu}$ and $G^a_{\mu \nu}$ are the field strengths of the SM gauge fields, while $\Lambda_{i}$ denote the effective scale of the non-renormalizable interaction. In the simplified model considered above, the condition $M_{\tilde{Q}} \gg M_S$ cannot hold for physically allowed parameters, see Fig.~\ref{fig:XSecPlot}. 
Thus the rates obtained by using the effective Lagrangian need to explicitly account for the loop function $A_{0}$ (i.e. non-trivial scaling of $\Lambda_{i}$) to get accurate results, even if the expansion $E/\Lambda_{i}$ naively seems to be well defined. 

Nevertheless, in this context the formalism of effective Lagrangian approach 
is very useful, since it allows to capture in a model-independent way the crucial information 
concerning the underlying dynamics responsible of generating the effective coupling. 
If the effective operator is generated perturbatively by integrating out particles running in the loops, its coefficient has the general form 
\begin{align}
\label{eff_scale}
	\frac{1}{\Lambda_{i}} = \frac{\alpha_{i}}{4\pi} \frac{N_{e} g_{\tilde{Q}S}}{m_{\tilde{Q}}} C_{i},
\end{align}
where $C_{i}$ is an $\mathcal{O}(1)$ factor originating from loop integrals and $g_{S}$ denotes an effective coupling between $S$ and the mediators and $N_{e}$ is the effective number of degrees of freedom running in the loops. 
The cross section obtained from the effective Lagrangian is roughly
\begin{equation}
	\sigma(pp \to S \to \gamma \gamma)
%	=	\frac{\pi^{2}}{8 M_{S}^3} I_{\rm pdf} \Gamma (S \to \gamma\gamma),
%	=	\frac{\pi^{2}}{8 M_{S}^3} \frac{\alpha^2 m_{S}^{3} N_e^{2} g_{\tilde{Q}S}^{2}}{64\, \pi^2\, m_{\tilde{Q}}^{2}} I_{\rm pdf} C^2_{\gamma \gamma}
	\approx	\frac{\alpha^2 N_e^{2}g_{S}^{2}}{512\, m_{\tilde{Q}}^{2}}.
\end{equation}
%where $I_{\rm pdf} \approx 5.8$ is the partonic integral, 
Fixing its value to  $5~\text{fb}$ suggests that in order to reproduce the required phenomenology of the observed diphoton excess, the effective coupling defined by Eq.~\eqref{eff_scale} should satisfy
\begin{align}
	 N_e g_{S} \approx 70 \times \frac{m_{\tilde{Q}}}{M_{S}}\, .
\end{align}
As we can see, this would require necessarily a $g_{S} \simeq \mathcal{O}(10)$
if $N_e\sim {\cal O}(1)$. Then, from these results
one can naively guess that this large number of $g_{S}$
points towards either strong dynamics or a relatively large number of degrees of freedom  in the loops. 
This is, indeed, justified conclusion if one considers vector-like fermions running in the loop~\cite{Franceschini:2015kwy,Knapen:2015dap,Pilaftsis:2015ycr,Buttazzo:2015txu,Angelescu:2015uiz,Gupta:2015zzs,Ellis:2015oso,McDermott:2015sck,Kobakhidze:2015ldh,Martinez:2015kmn,No:2015bsn,Chao:2015ttq,Fichet:2015vvy,Curtin:2015jcv,Falkowski:2015swt,Aloni:2015mxa}, where
 $g_S$ coincides with the corresponding fermion Yukawa coupling to the scalar resonance. In this respect, we qualitatively agree with the 
 effective model approach conclusions of Ref.~\cite{Aloni:2015mxa} on the large production rates.

However, as we have shown with  the present simplified model, when scalar fields are propagating in the loop, the above conclusions do not hold anymore. 
The coupling $g_S$ can be made naturally (and consistently) very large, even in the framework of weakly coupled field theories, 
being related to the ratio $g_{S} = \mu_{\tilde{Q}}/m_{\tilde{Q}} \sim\mathcal{O}(10)$. This is the advantage of having {\it soft} coupling $\mu_{\tilde{Q}}$
in theories with scalars. However, this requires that  the scalar resonance should not get a VEV or equivalently 
it should not be a Higgs-like particle. We have seen that constraints from the colour-charge breaking minima could limit the ratio $\mu_{\tilde{Q}}/m_{\tilde{Q}}$.
This implies that the present simplified model breaks down and Eq.~\eqref{eff_scale} does not correspond to the scale in Eq.~\eqref{Leff} any more.
The correct interpretation would require the knowledge of the full supersymmetric theory.

%%%%%%%%%%%%%%%%%%%%%%%%%%%%%%%%%%%%%%%%%%%%%%%%%%%%%%%%%%%%%%%%
\section{Discussion and Conclusions}

We have shown that the recently claimed evidence for the 750~GeV diphoton excess at the LHC
can actually, contrary to the general opinion, favour supersymmetry in Nature. However, the corresponding supersymmetric theory
must contain a singlet in addition to the SM particle content, and the mass spectrum of the sparticles must be rather unusual featuring
several light scalars while the gluino must be heavy to satisfy the LHC constraints. 

To study the diphoton excess we have presented a simplified model that captures the required properties of the supersymmetric
theory it is to be embedded in. As the result, we have shown that the NMSSM-like particle content is sufficient to generate large enough $gg\to S$ and $S\to \gamma\gamma$ processes at loop level to explain the observations. In particular, the coloured scalars in the loops have an advantage over the fermions to produce the 
needed large signal because of the possibly large dimensionful coupling $\mu_{\tilde{Q}}.$
We have also shown that the requirement of a colour and charge conserving vacuum constrains the parameter space of this scenario
so that the model is testable. In the context of the simplified model, that by itself is not supersymmetric, this implies that the model breaks
down at rather low energy where new superpartners of the complete supersymmetric model must appear to save physics.
The concrete prediction of our scenario is the existence of relatively light squarks which should be searched for
at the LHC. 

We conclude that, if this scenario will turn out to be the explanation of the diphoton excess, supersymmetry, indeed, was `just around the corner.'
However, to study the full the model and its precise properties would require more discoveries at the LHC or at the future 100~TeV collider.

%%%%%%%%%%%%%%%%%%%%%%%%%%%%%%%%%%%%%%%%%%%%%%%%%%%%%%%%%%%%%%%%
%\begin{acknowledgments}
%\end{acknowledgments}
\section*{Acknowledgments}
The authors thank Luca Marzola and Stefano Di Chiara for useful discussions. 
This work was supported by the grants IUT23-6, PUT716, PUT799 and by the EU through the ERDF CoE program.

%%%%%%%%%%%%%%%%%%%%%%%%%%%%%%%%%%%%%%%%%%%%%%%%%%%%%%%%%%%%%%%%

\section*{References}

\bibliographystyle{elsarticle-num}

\bibliography{750GeVdiphoton_v3}

\end{document}